\documentclass[final,1p,times]{elsarticle} 
\usepackage{graphicx} 
\usepackage{amssymb} 
\usepackage{amsthm} 
\usepackage{lineno} 

\journal{Nuclear Physics A} 
\begin{document} 

\begin{frontmatter} 


\title{Full Jet Reconstruction in Heavy Ion Collisions}

\author{Sevil Salur$^a$}

\address[a]{Lawrence Berkeley National Laboratory, 1 Cyclotron Road MS-70R0319, Berkeley, CA 94720}

\begin{abstract} 
Full jet reconstruction has traditionally been thought to be difficult in heavy ion events due to large multiplicity backgrounds. The search for new physics in high luminosity p+p collisions at the LHC similarly requires the precise measurement of jets over large backgrounds caused by pile up;  this has motivated the development of a new generation of jet reconstruction algorithms which are also applicable in the heavy ion environment. We review the latest results on jet-medium interactions as seen in A+A collisions at RHIC, focusing on the new techniques for full jet reconstruction. 
\end{abstract} 

\end{frontmatter} 



\section{Introduction}\label{intro}

Jets are remnants of hard-scattered quarks and gluons which are the fundamental objects of pQCD \cite{jetsref}.    They are 
well-defined objects that are measurable from the hadronic final-states and  calculable in pQCD from the partonic states \cite{seymor}.  
In high energy collisions of all kinds, they have been studied extensively to investigate the properties of quarks and gluons.  At RHIC, jets are put to a new use to study the hot QCD matter through their interaction and energy loss in the medium (``jet quenching") \cite{highpt}.

Until recently, to avoid the complex backgrounds of heavy ion events, inclusive hadron distributions and di-hadron correlations at high transverse momentum were utilized to measure jet quenching at RHIC indirectly.   However, these measurements of jet fragmentation particles are biased towards the population of jets that has the least interaction with the medium.  Models with a wide range of parameters are able to describe the measurements of di-hadron correlation and nuclear modification factors, leading to limited constraints upon the underlying physics \cite{bass,armesto,xin}.    Full exploration of jet quenching can be performed only when the geometric biases in measurements of A+A collisions are overcome. Jet reconstruction  at the partonic level with significantly reduced biases enables this study, with qualitatively new observables such as energy flow, jet substructure and fragmentation functions that can be measured in multiple channels (inclusive, di-jets, h-jets and gamma-jets).

The detector upgrades together with the increased beam luminosities of RHIC enable jet reconstruction in heavy ion collisions for the first time \cite{salurww}.  In this article we review the new techniques developed for full jet reconstruction in heavy ion collisions at RHIC and the results that help move our understanding forward. The experimental details of jet reconstruction measurements  utilizing the STAR and PHENIX experiments can be found in \cite{me, ploskonQM, yuiQM} for the inclusive spectra,  \cite{cainesQM,kapitanQM} for the underlying event, and  \cite{jor,brunaQM} for the accompanying jet fragmentation studies in heavy ion collisions.



\section{Jet Reconstruction Algorithms}

\begin{figure}[here!]
\centering
$\begin{array}{cc}
\resizebox{0.57\textwidth}{!}{
\includegraphics{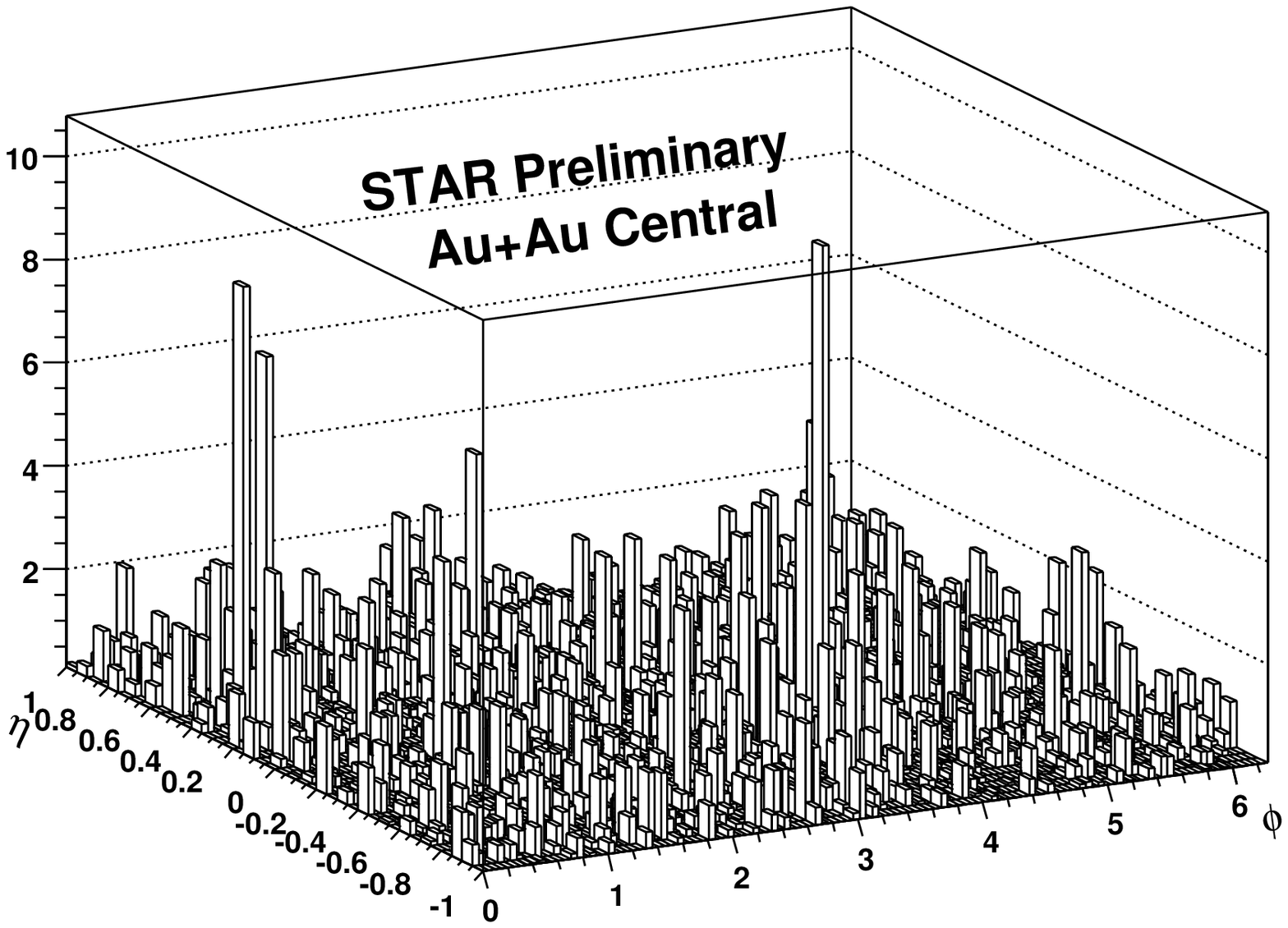}}&
\resizebox{0.40\textwidth}{!}{
\includegraphics{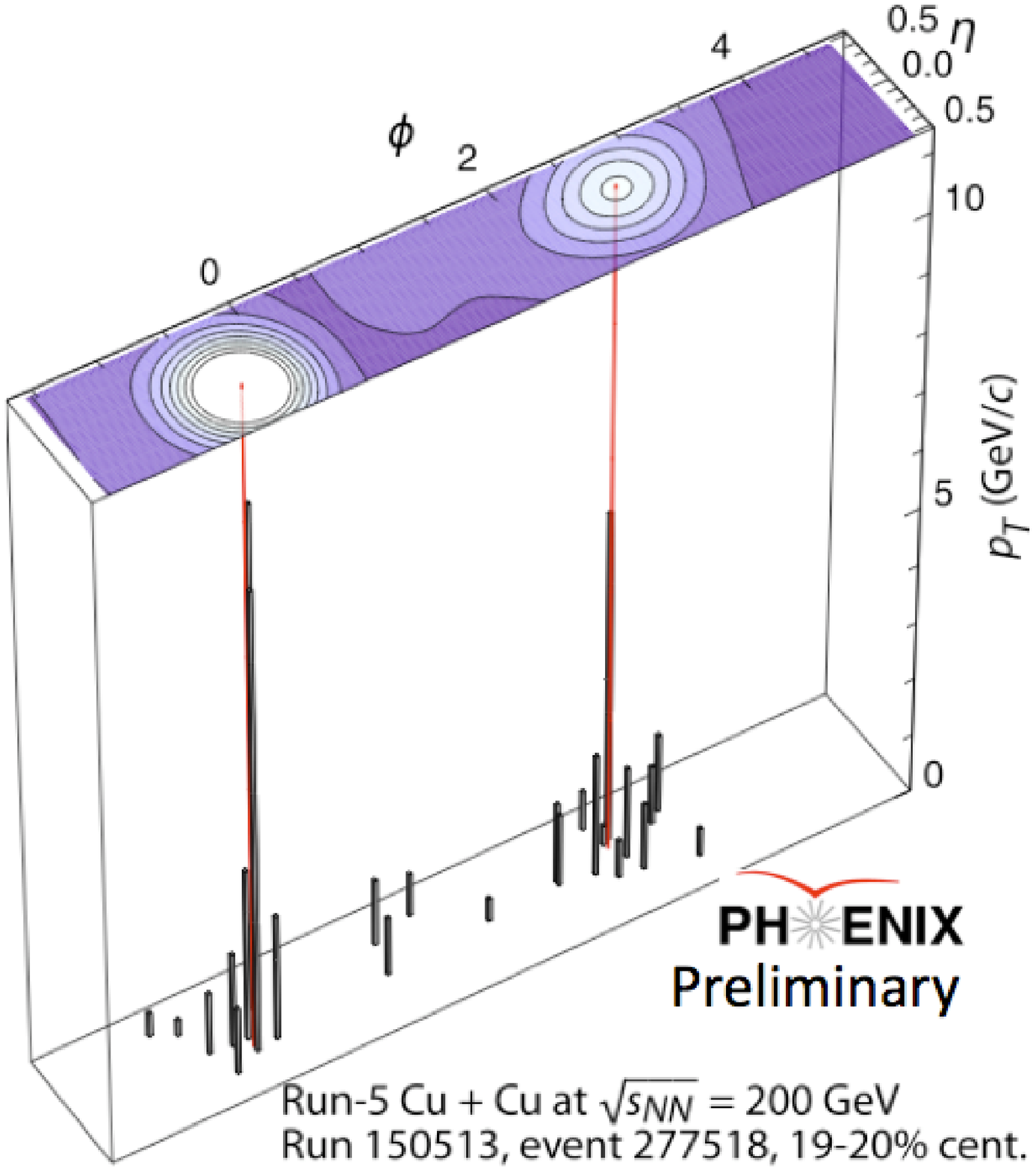}}
\\
\end{array}$

 \caption[]{Left panel: A reconstructed di-jet from a central Au+Au event at $\sqrt{s_{NN}}=200$ GeV in the STAR detector \cite{me,jor}. Right panel:  A reconstructed di-jet from a Cu+Cu event at  $\sqrt{s_{NN}}=200$ GeV in the PHENIX detector \cite{yuiQM}. 
   } \label{fig:dijets}
\end{figure}

During the last 20 years, various jet reconstruction algorithms have been developed for both leptonic and hadronic colliders.   For an overview of jet algorithms in high energy collisions, see  \cite{me,davidE} and the references therein.   Here we will briefly discuss the three algorithms (cone, sequential recombination and gaussian filtering) used for the RHIC analyses.  Figure~\ref{fig:dijets} shows an example of a di-jet event  for the central Au+Au collisions collected by the STAR experiment on the left panel and another di-jet event for the Cu+Cu collisions collected by the PHENIX experiment on the right panel.   For both cases clear peaks of di-jets can be observed over the heavy ion environments.

The cone algorithm is quite  intuitive and is based on the simple picture that a jet consists of a large amount of hadronic energy in a small angular region. Therefore, the main method for the cone algorithm is to combine particles in $\eta - \phi $ space with their neighbors within a cone of radius R ($\rm R=\sqrt{ \Delta \phi ^{2}+ \Delta \eta^{2} }$). To optimize the search and effectiveness of jet finding and  to accommodate higher-order processes, splitting, merging, and iteration steps can be used.  Unlike the cone algorithm, the sequential recombination algorithms  combine pairs of objects relative to the closeness of their $p_{T}$. Particles are merged into a new cluster via successive pair-wise recombination.  While the lowest $p_{T}$ particles are the starting point for clustering in the $k_{T}$ algorithm, in the anti-$k_{T}$ algorithm, recombination starts with the highest momentum particles.  Due to these different approaches, in the $k_{T}$ algorithm, arbitrarily shaped jets are allowed to follow the energy flow, resulting in less bias on the reconstructed jet shape than with the anti-$k_{T}$ or cone algorithm which are more or less restricted to a circular shape \cite{catchment}.  The Gaussian filtering type of algorithm simply extracts jets as local maxima in the $\eta-\phi$ space by linearly filtering particles.  Algorithmic details of cone,  sequential recombination and Gaussian filtering can be found in \cite{jets,kt,ktref,blazey, gauss, gauss1} and the references therein.

Motivated by the need for precision jet measurements in the search for new physics in high luminosity p+p collisions at the LHC, a new approach to jet reconstruction and background subtraction was developed  \cite{catchment,salamtalk}. A key feature in this procedure is a new QCD inspired algorithm for separating jets from the large backgrounds due to pile up. As it turns out from simulations,  these improved techniques can also be used in heavy ion environments where the background subtraction is essential for jet measurements.  Sequential recombination algorithms ($\rm k_{T}$, anti-$\rm k_{T}$ and Cambridge/Aachen (CAMB)) encoded in the $FastJet$ suite of programs  \cite{catchment,antikt}, along with a seedless infrared-safe cone algorithm (SISCone) \cite{sis} are utilized to search for jets in p+p \cite{elena} and Au+Au collisions collected by the STAR experiment.  Previously an alternative seeded cone algorithm  was also explored by STAR for Au+Au collisions in order to avoid instabilities in cone-finding due to large heavy ion background \cite{me,jor}.   The PHENIX experiment extracts jets successfully with a Gaussian filtering algorithm in p+p and Cu+Cu collisions \cite{yuiQM}.



\section{Heavy Ion Background}

Measuring jets above the complex heavy ion background is a challenging task. Full jet reconstruction in heavy ion collisions requires a fundamental assumption that the signal and the uniform background are two separable components.  However, this assumption can be violated by biases in the background estimation due to the presence of jets. Initial state radiation, even though expected to be small compared to jet energy, might be different in A+A relative to p+p. In the simpler p+p case, initial state processes resulting in the enhancement of the multiplicity of the underlying events appears to be distributed uniformly \cite{cainesQM}. Therefore it is fully accounted for the estimation of the background under jets \cite{catchment}. In Au+Au, ``the p+p correspondent" underlying event may be modified, possibly generating non-uniform structures. The non-uniformities in the background might be even larger  due to the final state processes.  Energy loss of the jet in matter might modify the event shape, resulting in non-uniform structures such as the ridge. Azimuthal and longitudinal anisotropy of heavy ion events will also result into non-uniform backgrounds.   Some of these sources of correlated backgrounds can be brought under quantitative control by using different collision systems. On the other hand, other observed effects might help us to understand details of the jet interactions with the heavy ion environment and may give further insight into the structures that are observed in di-hadron correlations and their origins.

With the assumption that the signal and the background are two separable components, the background correction can be estimated by following three steps.  The first step is measuring the jet area for the infrared safe algorithms. An active area of each jet is estimated by filling an event with  many very soft particles and then counting how many are clustered into a given jet. Fluctuations in the background also distort the jet spectrum towards larger  $E_{T}$ due to the steeply falling dependence of the jet production on $E_{T}$. This effect can be corrected  through an unfolding procedure (i.e., deconvolution).   
The second step is measuring the diffuse noise  (mean $p_{T}$ per unit area in the remainder of the event) and noise fluctuations.  The final step is correcting the jet energy by deconvolution of signal from the background, using parameters that are extracted from measurable quantities.


\section{Biases}

The ultimate goal of full jet reconstruction is to investigate the jet quenching in heavy ion collisions at the partonic level, without any ambiguities being introduced by hadronization and geometric biases of the inclusive spectrum and di-hadron measurements.  However,  it is possible that  new biases can be introduced when reconstructing jets. For example, all jet algorithms have various parameters for searching and defining jets, and the effects of varying these parameters need to be explored  in detail for a full understanding of jet reconstruction.

\begin{figure}[h!]

\centering
\resizebox{1.\textwidth}{!}{
\includegraphics{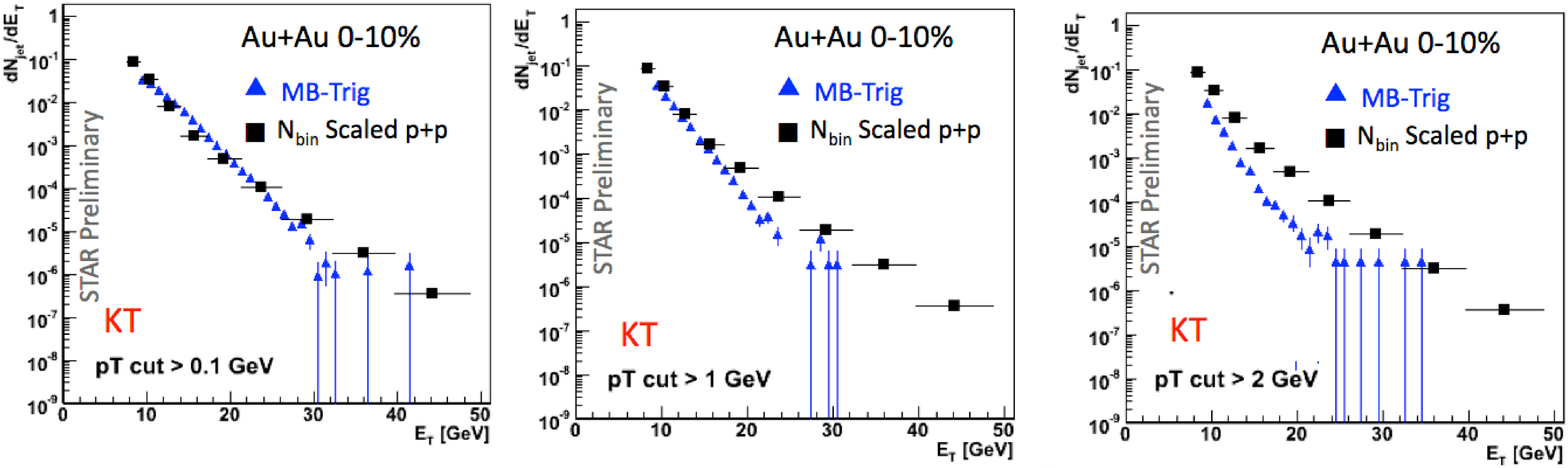}}
\\
 \caption[]{The $\rm p_{T}$ threshold dependent comparison of inclusive jet spectra in heavy ion and p+p collisions.
The filled triangle symbols are from MB-Trig, and filled squares are from $\rm N_{Binary}$ scaled p+p collisions.
In addition to $\rm p_{T}$ cuts shown in the figures, R is set to 0.4. \cite{me}.
 
   } \label{fig:ptcut}
\end{figure}

A bias will be introduced while trying to reduce the effect of the background fluctuations in heavy ion collisions with the threshold cuts on the track momenta and calorimeter tower energies ($p_{T}^{cut}$). Figure~\ref{fig:ptcut} shows the comparison of the jet spectra reconstructed by the $k_{T}$ algorithm with a variation of $p_{T}$ threshold cuts in Au+Au and the $\rm N_{Binary}$ scaled p+p collisions \cite{me}. 
For these results, energy resolution of the detectors and the underlying heavy ion background fluctuations were corrected with multiplicative factors of the jet spectrum estimated  by utilizing  Monte-Carlo model studies based on Pythia 8.107 \cite{pythia}.  While the agreement between Au+Au and $\rm N_{Binary}$ scaled p+p  jet measurements is good for the lowest value of the $p_{T}$ cut, it is also seen to be poorer with the larger
$p_{T}$ threshold cut. This suggests that the threshold cuts introduce biases which are not fully corrected by the procedures that use fragmentation models that are developed for $\rm e^{+} + e^{-}$ and p+p collisions. 

To enhance the recorded rate of high $p_{T}$ particles and jets, events above some threshold in the electromagnetic calorimeter  are collected. (This threshold is 5.4 GeV for the STAR experiment during data taking in years 2006 and 2007.)   This is very similar to the case of jets that are reconstructed with seeded infrared unsafe algorithms.  
The online calorimeter tower triggers introduce a strong bias of reconstructed jets that are fragmenting hard in comparison to the jets that are reconstructed without a seed. 

The resolution parameter or cone size, which restricts the area of the jet and thereby the amount of energy flow, can be a harder parameter to calculate hence interpret in heavy ion collisions than in p+p collisions.  If the jets are broader  in the heavy ion environment, the same resolution parameter might not be sufficient to recover the same fraction of jet energy in comparison to p+p jets.  This bias needs to be investigated by varying the resolution parameter and by looking into the jet profile of these jet definitions.


\section{Results}

\begin{figure}[h!]

\centering
$\begin{array}{cc}
\resizebox{0.47\textwidth}{!}{
\includegraphics{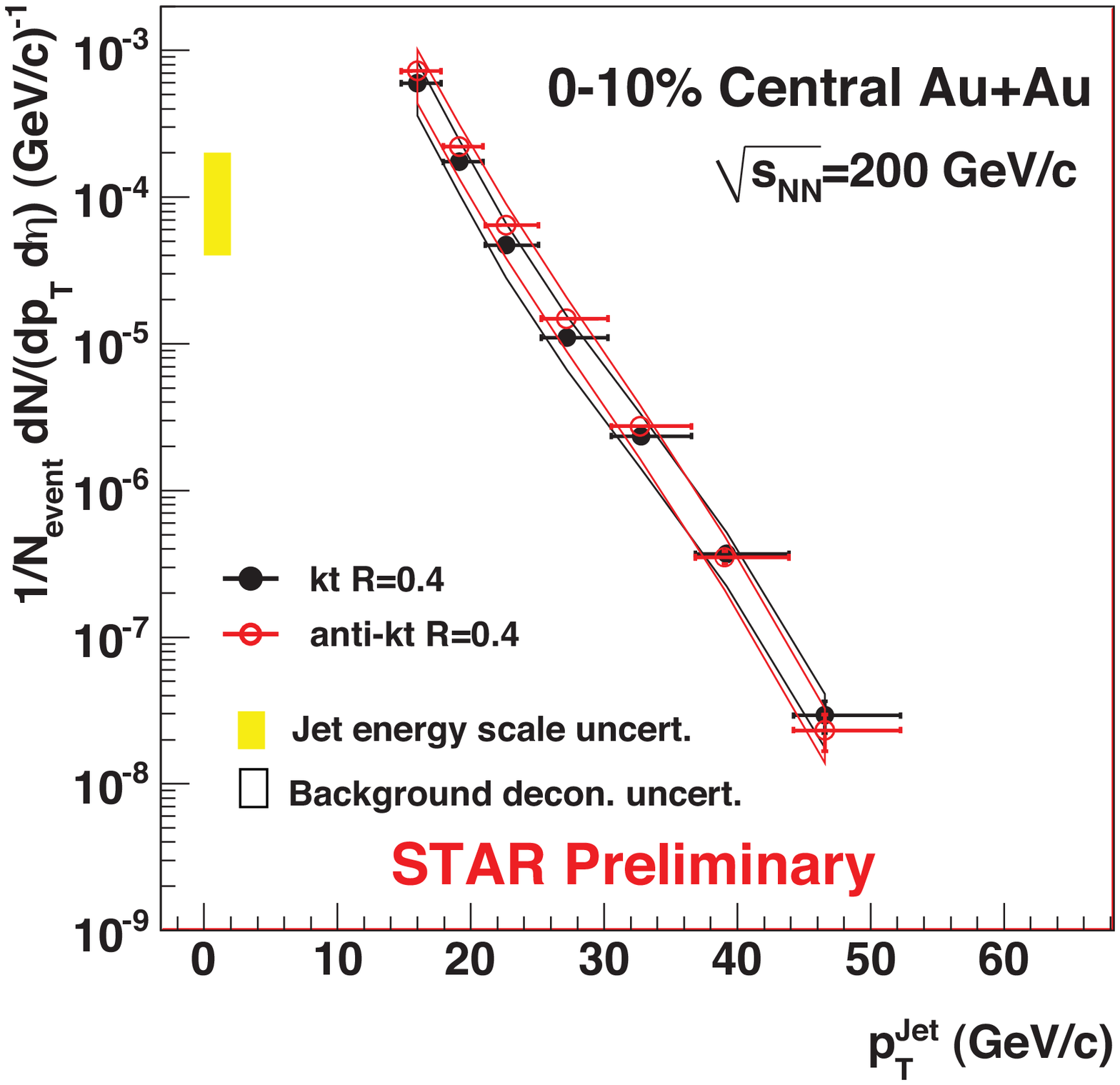}}&
\resizebox{0.53\textwidth}{!}{
\includegraphics{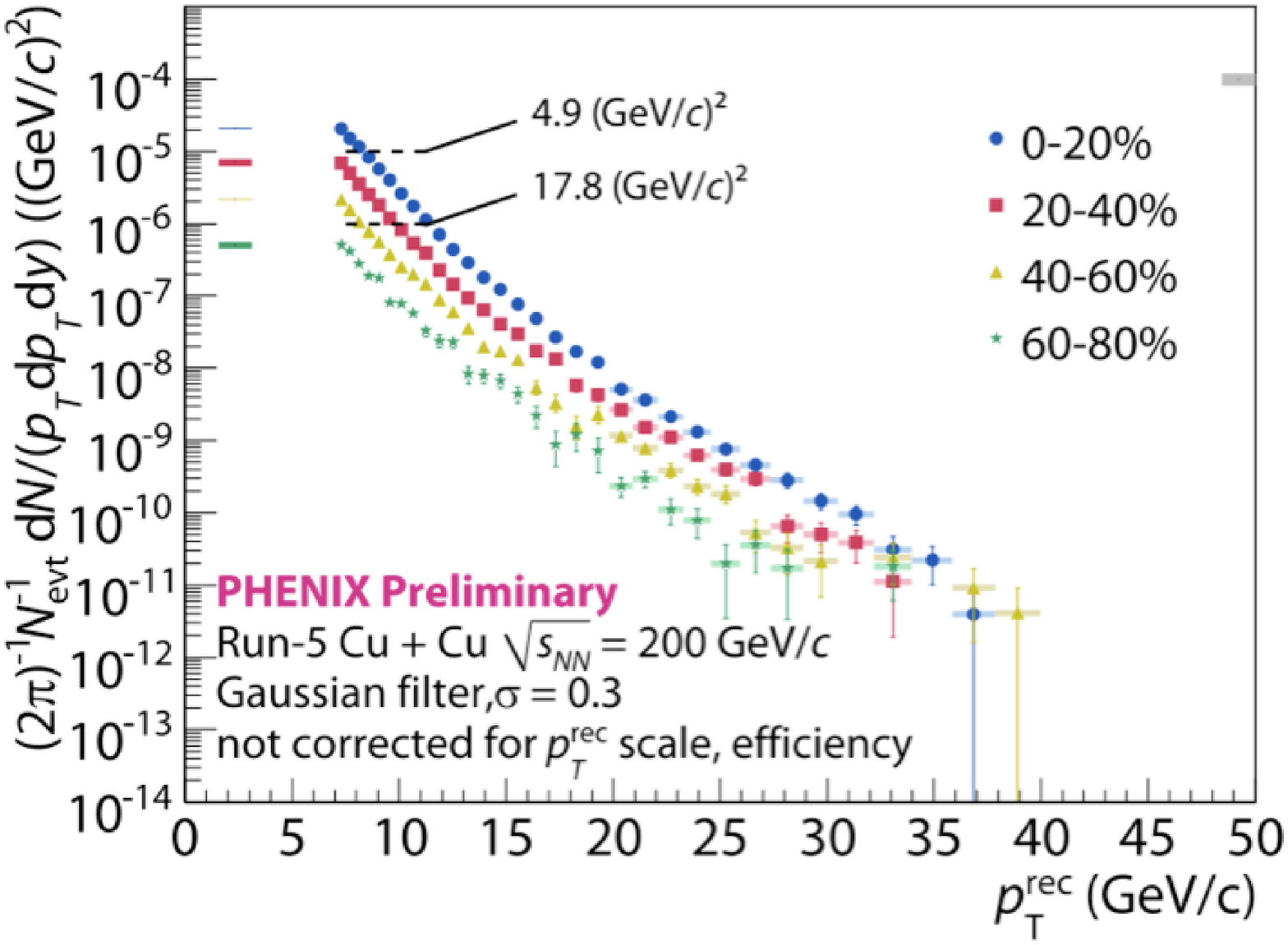}}
\\
\end{array}$

 \caption[]{Left panel: Jet yield per event vs transverse jet energy ($E_{T}$) for the central Au+Au collisions obtained by the sequential recombination ($\rm k_{T}$ and anti-$\rm k_{T}$) algorithms from STAR  \cite{ploskonQM}. Right panel: Uncorrected jet yield vs jet $E_{T}$  for the Cu+Cu collisions obtained by the gaussian filtering algorithm on the right \cite{yuiQM}.  See insets for the fixed parameters of the jet reconstruction algorithms. } \label{fig:kt}
\end{figure}

The left panel of Figure~\ref{fig:kt} shows the comparison of the inclusive jet spectra reconstructed by sequential recombination algorithms for central Au+Au collisions collected by the STAR experiment \cite{ploskonQM}.  Systematic uncertainties due to the unfolding procedure  are shown as the envelopes in red and black and the jet energy resolution as the yellow bar.  Jet spectra are consistent between the two  sequential reconstruction algorithms extending to 50 GeV kinematic reach.  Uncorrected jet spectra reconstructed with a Gaussian filtering algorithm for various Cu+Cu centralities collected by the PHENIX experiment are presented in the right panel of Figure~\ref{fig:kt}.  Within the restricted experimental acceptance, a large $p_{T}$ range of jets can also be reconstructed \cite{yuiQM}.  The same algorithms that are used for heavy ions are also used to reconstruct jets in p+p collisions for both experiments. The jet spectra reconstructed in p+p collisions from various algorithms with the fixed parameters (sigma of 0.3 for gaussian filtering and the resolution parameter of 0.4 for the sequential recombination algorithm) all agree well with the previously published RHIC results using a cone algorithm with split merge steps (cone radius of 0.4) \cite{ploskonQM,yuiQM,starpp,nlo}.


\begin{figure}[h!]

\centering
$\begin{array}{cc}
\resizebox{0.42\textwidth}{!}{
\includegraphics{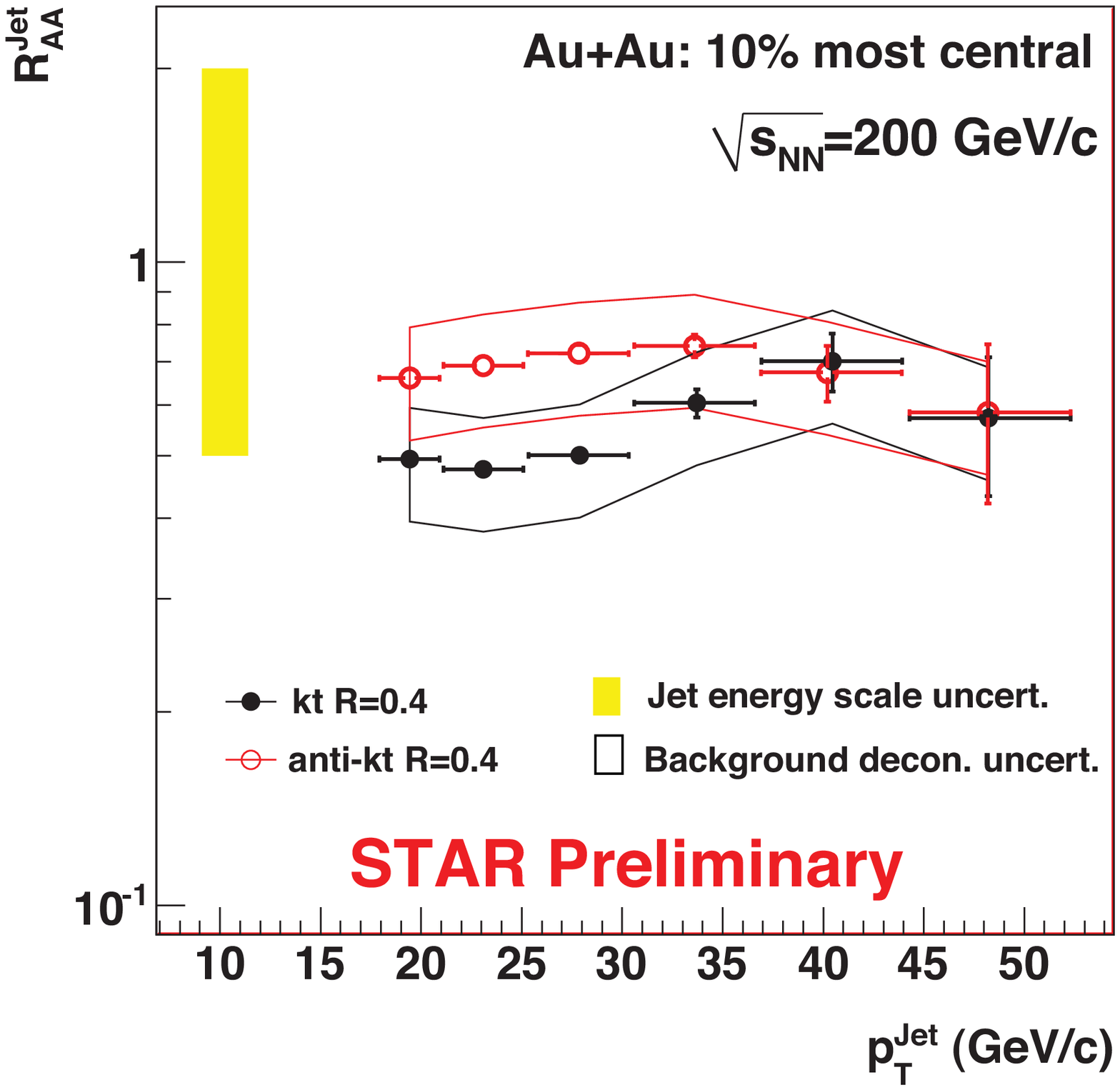}} &
\resizebox{0.54\textwidth}{!}{
\includegraphics{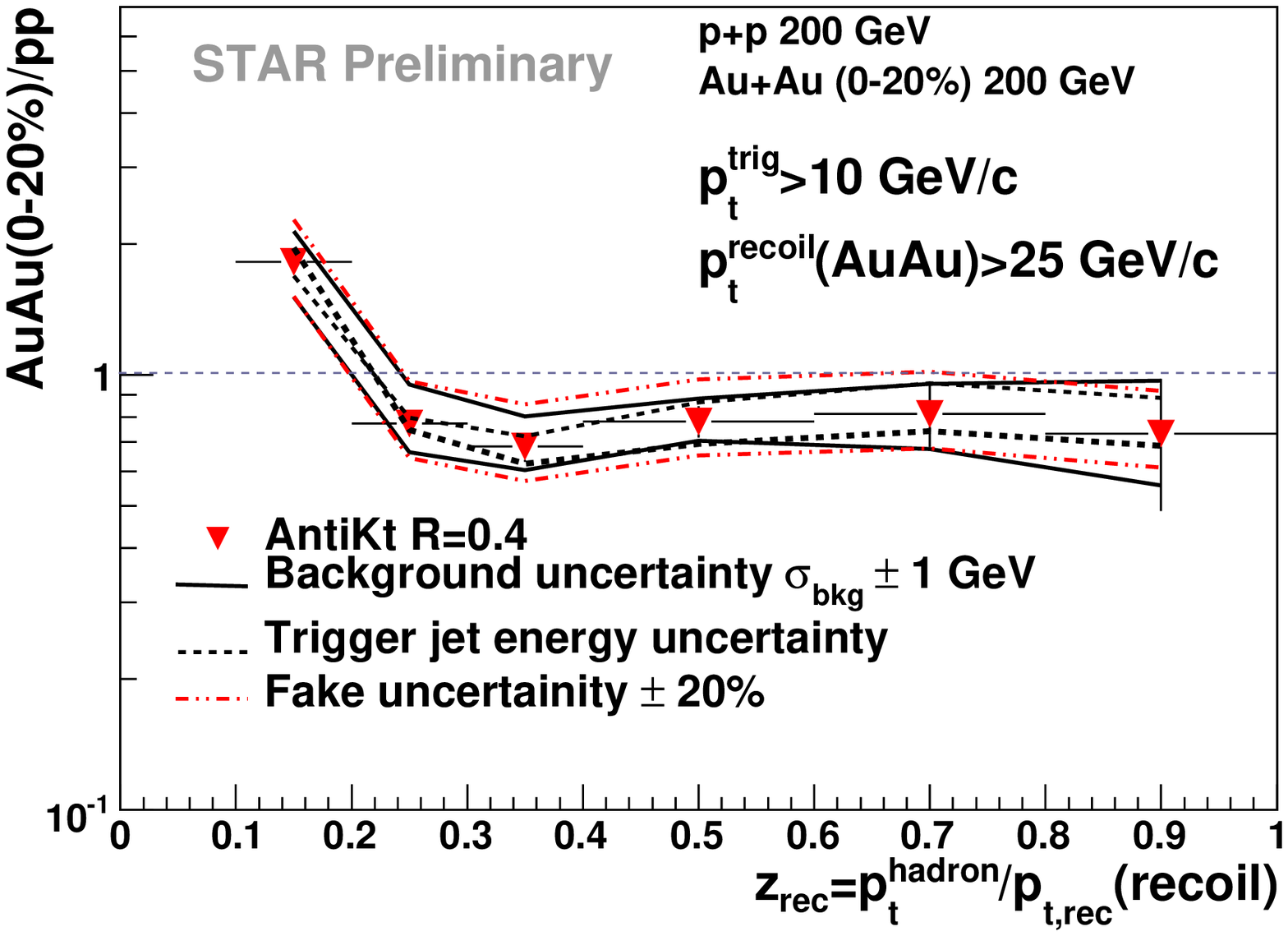}}
\\
\end{array}$

 \caption[]{Left panel: Momentum dependence of the nuclear modification factor of jet spectra reconstructed with $\rm k_{T}$ and anti-$\rm k_{T}$ algorithms (0-10\% most central Au+Au divided by $N_{bin}$ scaled p+p collisions) \cite{ploskonQM}. Right panel:  $z$ dependence of the ratio of the fragmentation functions for recoil jets (normalized to the number of recoil jets) 
measured in 0-20\% central High Tower triggered Au+Au events to High Tower triggered p+p collisions \cite{brunaQM}.  The systematic uncertainties in the estimation of the width of the Gaussian parameterization of the background fluctuations are presented as solid envelopes. 
 
   } \label{fig:raa}
\end{figure}

 The nuclear modification factor for the reconstructed jet spectra with the resolution parameter of 0.4 from $\rm k_{T}$  and anti-$\rm k_{T}$ can be found in the left panel of Figure~\ref{fig:raa} \cite{ploskonQM}.  The envelopes shown represent the one sigma uncertainty of the deconvolution of the heavy ion background. The total systematic uncertainty on the jet energy scale is around 50\%, shown as the yellow bar.  In the case of full jet reconstruction, $\rm N_{Binary}$ scaling as calculated by a Glauber model \cite{glauber} is expected if the reconstruction is unbiased, i.e. if the jet energy is recovered fully independent of the fragmentation details, even in the presence of strong jet quenching. This scaling is analogous to the cross section scaling of high $p_{T}$ direct photon production in heavy ion collisions observed by the PHENIX experiment \cite{phenix}.   A large fraction of jets are reconstructed 
when using both $k_{T}$ and anti-$k_{T}$ sequential recombination algorithms with a resolution parameter of 0.4. Momentum dependence of the nuclear modification factor is also different than the observed suppression of the $\pi$ meson $\rm R_{AA}$. However even though there are large systematic uncertainties, a hint of a suppression of jet $\rm R_{AA}$ above 30 GeV can be observed. This implies that an unbiased jet reconstruction is not reached fully.  It is expected that for a smaller resolution parameter, this suppression should reach to single particle suppression at large momentum. For the case of R=0.2, further suppression is observed to grow to larger values supporting 
this expectation \cite{ploskonQM}.

Medium effects at parton splitting can also be studied with fragmentation functions. Fragmentation function measurements are extremely difficult due to the uncertainty in the statistical subtraction of the background particles and to potential biases in the reconstruction discussed earlier.  Nevertheless, the right panel of Figure~\ref{fig:raa} is a first attempt of the $z$ ($z=p_{T}^{hadron}/p_{T}^{Jet}$) dependence of the ratio of the fragmentation functions from the recoil jets of the di-jet coincidence measurements in 0-20\% central Au+Au to p+p collisions \cite{brunaQM}. See the inset of the right panel in Figure~\ref{fig:raa} for the applied selection cuts for the trigger and the recoil jets reconstructed with the anti-$k_{T}$ algorithm. Various uncorrelated systematic uncertainties are also included as the solid and the dotted histograms.  Within the given systematic uncertainties, the ratio of the fragmentation functions of Au+Au to p+p show no significant suppression. This is surprising given the expectation of softening due to jet quenching. The lack of modification in fragmentation functions implies that the di-jets that are reconstructed in this analysis are biased towards a sample of unquenched jets that are only coming from the surface.  It is also possible that due to the broadening of the jets and the requirement of the limited resolution parameter only part of the jet energy is recovered. This results in an insensitivity to expected softening when measuring fragmentation functions statistically.


 This is the first time that these large momenta reaching 50 GeV can be studied in heavy ion collisions with the expanded kinematic reach of full jet reconstruction.  New physics effects should be considered when interpreting the results at large momentum.  An example is the momentum dependence of the relative contributions of quark and gluon sub-processes to inclusive jet production.  It is  known that  in elementary p+p collisions these contributions vary with respect to the jet momentum \cite{vogelsang}.  In a heavy ion environment when quark gluon plasma is produced,   the overall effect of these relative contributions might modify the  expected shape of the spectra and therefore of the nuclear modification factors. Another possibility is that at large momentum fraction $x$, initial state effects (such as the EMC effect which is the deviation between structure functions of heavy ions to light ions) are observed to be as large as 15\% \cite{emc}.  There might be other effects playing a major role in the relative suppression or enhancement of nuclear modification factors at large momentum.  

 \section{Discussion}
 
 Quantitative analysis of jet quenching in heavy ion collisions requires model building.  The new Monte-Carlo based simulations of  jet quenching in medium such as Jewel \cite{jewel}, Q-Pythia \cite{qpythia} and YaJEM \cite{yajem} and complementary analytic  calculations \cite{vitev,vitev2,borghini}  recently became  available.  A strong broadening of showers in transverse momentum with respect to the jet axis results in wider angular distributions due to quenching in these analytic calculations.  However,  Monte-Carlo simulations such as Jewel observe no significant broadening when $p_{T}$ threshold cuts are applied \cite{jewel}.  As observed experimentally and discussed earlier, applying momentum cuts to reduce the background fluctuations introduces biases and limits the broadening observable.  There are also many uncertainties (e.g., how hadronization is treated) in the predictions  of these models and calculations.  To confront the calculations with data, new robust QCD jet observables that are unaffected by the $p_{T}$ cuts and hadronization need to be explored experimentally. The  subjet observable is infrared safe and insensitive to hadronization and may be used to study the jet quenching \cite{jewel}.  

Unbiased reconstruction of jets in central heavy ion collisions at RHIC energies would be a breakthrough to investigate the properties of the matter produced at RHIC. 
New theoretical developments  such as  the FastJet suite of jet reconstruction algorithms and new medium-modified shower MC codes such as Q-Pythia, JEWEL, YaJEM and others enable study of jet quenching at the partonic level. The studies shown here indicate that reconstruction of jets may indeed be possible in heavy ion events. Currently a uniquely large kinematic limit  is reached (up to 50 GeV) with full jet reconstruction at RHIC.   
Jet reconstruction in heavy ion collisions is not yet free of geometric biases.  Biases are introduced due to selection of particles such as $p_{T}$ cuts to reduce the fluctuations of heavy ion background, requirement of algorithmic parameters such as cone size or resolution parameter, and the collection of events with thresholds to enhance the jet rates. To assess fully the systematic uncertainties of jet measurements these effects must be investigated further. 
These results motivate us to look further into multiple channels for consistency checks (inclusive, di-jets, h-jets, gamma-jets to measure qualitatively new observables: energy flow, jet substructure, fragmentation function \cite{vitev,vitev2}.)

A copious production of very energetic jets, well above the heavy ion background, is predicted to occur at the LHC \cite{peter}. The large kinematic reach of high luminosity running at RHIC and at the LHC may provide a sufficient lever-arm to map out the QCD evolution  of jet quenching \cite{solan}.  The comparison of  full jet measurements in the different physical systems generated at RHIC and the LHC will provide unique and crucial insights into our understanding of jet quenching and the nature of hot QCD matter.



\section*{Acknowledgments}
The author wishes to thank the organizers of 2009 Quark Matter for the fruitful meeting, and  the PHENIX and STAR Collaborations for providing the data presented here.

\end{document}